\documentclass[12pt]{article}
\usepackage[T1]{fontenc}
\usepackage{setspace}
\usepackage{amsmath,amssymb,amsthm,mathtools}
\usepackage{lmodern}
\usepackage[svgnames]{xcolor}
\usepackage[round]{natbib}
\usepackage{titlesec}
\usepackage{enumitem}
\usepackage[margin=1.25in]{geometry}
\usepackage{hyperref}
\hypersetup{
   pdftitle={Testing Experts Who Can Learn},
    pdfauthor={Mark Whitmeyer and Kun Zhang},
    colorlinks=true,
    citecolor=DarkBlue,
    bookmarksnumbered=true,
    urlcolor=Indigo,linkcolor=DarkBlue
}
\usepackage[noabbrev,capitalise,nameinlink]{cleveref}

\titleformat{\section}{\normalfont\Large\bfseries\color{DarkRed}}{\thesection}{1em}{}
\titleformat{\subsection}{\normalfont\large\bfseries\color{DarkRed}}{\thesubsection}{1em}{}

\DeclareMathOperator{\conv}{co}

\newtheorem{theorem}{Theorem}
\newtheorem{proposition}{Proposition}
\theoremstyle{definition}
\newtheorem*{definition}{Definition}
\newtheorem{assumption}{Assumption}
\newenvironment{altassumption}[1]
{\renewcommand\theassumption{#1}\assumption}
  {\endassumption}
\crefname{assumption}{Assumption}{Assumptions}
\newcommand{\dd}{\,\mathrm{d}}
\begin{document}
\author{Mark Whitmeyer\thanks{Arizona State University. Email: \href{mailto:mark.whitmeyer@gmail.com}{mark.whitmeyer@gmail.com}.} \and Kun Zhang\thanks{Auburn University and University of Queensland. Email: \href{mailto:kun@kunzhang.org}{kun@kunzhang.org}.}}

\title{Testing Experts Who Can Learn\thanks{This paper supersedes an earlier version circulated under the title ``Redeeming Falsifiability?'' We thank the Editor and four anonymous reviewers for their excellent editorial comments, which greatly improved the paper. We are also grateful to Ashwin Kambhampati, Yonggyun Kim, Andreas Kleiner, Jonathan Libgober, Joseph Whitmeyer, and Renkun Yang for helpful comments and discussions. We thank participants at various conferences for their feedback. All remaining errors are our own.}}
\date{\today}

\maketitle

\begin{abstract}
We show that when expertise lies not in what one initially knows but in the ability to acquire information and thereby refine beliefs, prediction-based contracts can distinguish experts from non-experts.
\end{abstract}

\newpage

\section{Introduction}\label{introduction}
What distinguishes an expert from a charlatan? Much of the literature on testing strategic forecasters assumes that an expert knows the probabilities governing future outcomes, whereas a charlatan does not. We study a different conception of expertise: it is the ability to acquire information and thereby refine one's belief. That is to say, rather than ``knowing the answers,'' it is ``knowing where (or how) to look.'' Evidence from actual forecasting practice suggests that this distinction matters.\footnote{In geopolitical forecasting tournaments, forecasters who revised their predictions more often accessed more information and became more accurate over time, even though how often they revised was unrelated to the accuracy of their initial forecasts \citep{atanasov2020small}; among professional inflation forecasters, a contest rewarding accuracy induced more frequent revisions and more accurate forecasts \citep{gaglianone2022incentive}. See also \citet{mellers2015identifying} on the information acquisition of high-performing geopolitical forecasters.} Moreover, this notion of expertise is natural when advice is needed before all relevant evidence has been collected. Two advisers may, for example, assign the same probabilities to the outcomes of a clinical trial, yet only a specialist knows how to design and interpret a separate pilot study that distinguishes competing models of the treatment's efficacy.

In our model, a principal, Alice, offers Bob, a self-proclaimed expert, a contract whose payments may depend on his report and the realized outcome. Alice cannot tell whether Bob is an expert or a charlatan and does not know what he believes about the probabilities governing the outcome. After accepting, an expert may acquire information before reporting; a charlatan cannot. We say that a contract \textit{screens charlatans} if acceptance is optimal over the entire set of expert beliefs, while a charlatan strictly prefers rejection for at least one belief. \textit{Strong screening} retains the universal-acceptance requirement for experts but requires strict rejection regardless of the charlatan's beliefs.

When the expert cannot acquire information, any contract acceptable at every expert belief is also acceptable to a charlatan no matter what he believes.\footnote{Earlier work poses the same question by asking whether an uninformed forecaster can randomize over reports to secure a nonnegative payoff under every model \citep{olszewski2007contracts,babaioff2011valuable}. By the minimax theorem, such a report choice exists exactly when acceptance is optimal under every charlatan belief.}
Moreover, even with information acquisition, no contract can strongly screen charlatans. Intuitively, every belief an expert can hold before or after acquiring information is also a possible belief of the charlatan; if the charlatan strictly rejects at every such belief, acquiring information cannot make acceptance optimal for the expert.

We establish screening via a simple contract studied by \citet{falsifiability}: Bob receives a payment, names the outcome he considers least likely, and pays a penalty if that outcome occurs. Information benefits the expert by reducing his probability of paying the penalty. At the expert beliefs for which acceptance is hardest to secure, we assume that this reduction is at least a known amount and its cost is at most a known bound.
Because the expert's benefit from acquiring information is proportional to the penalty, these bounds allow Alice to choose a penalty large enough for that benefit to exceed the cost. She can then choose the payment so that the contract screens charlatans. At some belief, a charlatan strictly prefers rejection because he cannot reduce the risk of paying the penalty, while an expert with that belief can acquire information and make acceptance worthwhile.

\subsection{Related Literature}\label{related-literature}

This paper belongs to the literature on testing strategic forecasters. A central question in that literature is whether a forecaster who does not know the probabilities governing future outcomes can nevertheless choose reports that pass a test designed to ascertain whether he has that knowledge. Calibration \citep{dawid1982well} illustrates the difficulty: it is a foundational standard for evaluating forecasts, but can often be achieved without knowledge of the governing probabilities.\footnote{See \citet{olssurvey} for a survey. On calibrated or otherwise successful forecasting without knowledge of the governing probabilities, see \citet{asscal}, \citet{foster1999proof}, \cite{sandroni2003calibration}, \citet{hart2022calibrated}, and \citet{olszewski2007contracts, olszewski2008manipulability,olszewski2009manipulability}. For works exploring how informed and uninformed forecasters can be distinguished, refer to \citet{dekel2006non}, \citet{olszewski2009nonmanipulable}, \citet{hu2013expressible}, and \citet{pomatto2021testable}.} We replace passage of a test with willingness to accept an outcome-contingent contract.

The papers closest to ours ask whether every informed forecaster can be induced to accept a contract that an uninformed forecaster rejects. \citet{olszewski2007contracts} establish an impossibility when expertise consists of knowing which model is correct, whose logic our no-learning result restates for our environment (\cref{prop:general-limits}\ref{prop1-nolearning}). \citet{falsifiability} establish a related failure for simple outcome-contingent contracts. In our positive result (\cref{main}), we illustrate that information acquisition raises the expert's payoff at the beliefs where acceptance is hardest to secure, making expert acceptance compatible with rejection by charlatans.\footnote{Similar positive results obtained by restricting the possible outcome distributions or forecasting rules include \citet{olszewski2009strategic} and \citet{babaioff2011valuable}, which we discuss following \cref{main}. Related papers pursue useful objectives without requiring the tester to identify whether the forecaster is informed: \citet{echenique2007harm} protect the decision maker from harmful incorrect models, while \citet{olszewski2011principal} study dynamic contracts that attain the first-best payoff.}
Learning also plays a role in \citet{alnajjar2010testing} and \citet{deb2018evaluating}. In the former, observations accumulate until an informed forecaster can make a sufficiently precise prediction. The latter examines how to evaluate forecasters who receive information of differing precision over time.

Finally, there is a literature studying how to induce an agent to acquire information and (honestly) report what he finds.\footnote{Refer to, among others, \citet{zermeno2011}, \citet{rappoport2017incentivizing}, \citet{clark2021contracts}, \citet{yoder2022designing}, \citet{whitmeyer2023buying}, \citet{sharma2024procuring}, and \citet{li2026forecasters}. \citet{muller} formulate an ``ignorance equivalent,'' which they apply to the contracting-for-information-acquisition problem.}
We instead ask whether Alice can detect the ability to learn itself from his willingness to accept.

\section{Model and Results}\label{model}

There is a finite set \(\Omega=\{\omega_1,\ldots,\omega_n\}\) of possible outcomes, where \(n\geq2\). For any \(\nu\in\Delta(\Omega)\), write \(\nu_i\coloneqq\nu(\omega_i)\). Alice contracts with a self-proclaimed expert, Bob. Bob is risk neutral, and his payoff from rejecting any contract is zero.

A model \(\tau\in T\subseteq\Delta(\Omega)\) is a probability distribution over outcomes, where \(T\) is a
closed set containing at least two models.
The set \(T\) is common knowledge.
Bob is either an expert or a charlatan. If Bob is an expert, he holds a prior probability distribution over models \(Q\in\Delta(T)\), where \(\Delta(T)\) is endowed with the weak topology. This prior induces a distribution over outcomes
\[
\mu(Q)\coloneqq\int_T\tau \dd Q(\tau),
\]
which we write simply as \(\mu\) when no confusion can arise.
We allow any prior \(Q\) that is not concentrated on a single model, which we term nondegenerate.

An expert may acquire additional information about the model.
An experiment is a pair \(E=(S,\chi)\), where \(S\) is a measurable space of possible signals and \(\chi \colon T \to \Delta(S)\) is a measurable stochastic map. Conditional on \(\tau\), the signal is drawn according to \(\chi(\cdot\mid\tau)\), independently of the outcome, which is drawn according to \(\tau\). Consequently, observing the signal changes the expert's prediction of the outcome only by changing his posterior over models. Let \(\mathcal E\) denote the experiments available to expert Bob. Given \(Q\) and \(E\), let \(\overline\chi_Q\) denote the induced distribution of signals, let \(Q^s\) denote Bob's posterior over models after signal \(s\), and let \(\mu^s=\int_T\tau \dd Q^s(\tau)\) denote the induced posterior distribution over outcomes. The cost of acquiring information is given by a cost function \(C\colon \mathcal E\times\Delta(T)\to\mathbb R_+\cup\{\infty\}\), where \(C(E,Q)\) is the cost of conducting experiment \(E\) for an expert with prior \(Q\).

A charlatan cannot acquire information and may hold any prior \(P\in\Delta(T)\), inducing the distribution \(\rho(P)\coloneqq\int_T\tau \dd P(\tau)\) over outcomes. Alice does not observe whether Bob is an expert or a charlatan, nor does she observe his prior over models.

A contract is \(G=(M,t)\), where \(M\) is a nonempty compact metric message space and \(t\colon M\times\Omega\to\mathbb R\) is continuous. If Bob accepts, he chooses a message \(m\in M\), possibly at random. If outcome \(\omega_i\) occurs, Bob's net transfer is \(t(m,\omega_i)\). Let
\[
X_G\coloneqq\conv\left\{\left(t(m,\omega_1),\ldots,t(m,\omega_n)\right)\colon m\in M\right\},
\]
where \(\conv\) denotes convex hull. The set \(X_G\) is the compact convex set of expected payoff vectors attainable by possibly mixed messages. For a distribution over outcomes \(\nu\in\Delta(\Omega)\), Bob's value from reporting optimally under \(G\) is
\[
v_G(\nu)\coloneqq\max_{x\in X_G}\nu\cdot x.
\]
An expert with prior \(Q\) can report without acquiring information and obtain \(v_G(\mu(Q))\), or conduct an experiment \(E\in\mathcal E\) and obtain
\[
\int_S v_G(\mu^s) \dd\overline\chi_Q(s)-C(E,Q).
\]
A charlatan cannot acquire information, so under prior \(P\) his value from accepting \(G\) is \(v_G(\rho(P))\).

The timing is as follows. First, Alice offers a contract \(G\). Second, Bob accepts or rejects. Third, an expert who accepts may privately choose an experiment and observe its signal; a charlatan cannot acquire information. Fourth, Bob chooses a message to report. Finally, the outcome is publicly observed, and Bob receives \(t(m,\omega_i)\).

\begin{definition}
A contract \(G\) \emph{screens charlatans} if every expert can obtain a nonnegative payoff from accepting and \(v_G(\rho(P))<0\) for at least one \(P\in\Delta(T)\). It \emph{strongly screens charlatans} if every expert can obtain a nonnegative payoff from accepting and \(v_G(\rho(P))<0\) for every \(P\in\Delta(T)\).
\end{definition}

Both notions require acceptance to be optimal for every expert, differing only in what is required of the charlatan: screening permits him to accept under some priors, whereas strong screening requires him to strictly prefer rejection under every prior.

\subsection{Limits of Screening}

For a contract \(G\), define the charlatan's strict-rejection set as the set of priors under which he strictly prefers rejection to acceptance:
\begin{equation}\label{eq:general-rejection}
\mathcal R_G\coloneqq\left\{P\in\Delta(T):v_G(\rho(P))<0\right\}
=\left\{P\in\Delta(T)\colon \rho(P)\cdot x<0\text{ for every }x\in X_G\right\}.
\end{equation}
This set \(\mathcal R_G\) is convex and open in \(\Delta(T)\).

\begin{proposition}\label{prop:general-limits}~
\begin{enumerate}[label={(\roman*)},itemsep=1pt,topsep=3pt]
\item Without learning, no contract screens charlatans. If acceptance is optimal for every expert, it is optimal for a charlatan under every prior. \label{prop1-nolearning}
\item No contract strongly screens charlatans. \label{prop1-strong}
\end{enumerate}
\end{proposition}

\begin{proof}
For part \ref{prop1-nolearning}, without learning an expert with prior \(Q\) obtains \(v_G(\mu(Q))\). The distributions \(\mu(Q)\) induced by expert priors are dense in \(\conv(T)\). Since \(v_G\) is continuous, if acceptance is optimal for every expert, then
\[
v_G(\nu)\geq0\qquad\text{for every }\nu\in\conv(T).
\]
Because \(\rho(P)\in\conv(T)\), acceptance is optimal for a charlatan under every prior \(P\).

For part \ref{prop1-strong}, suppose the charlatan strictly rejects under every prior. Since \(\{\rho(P)\colon P\in\Delta(T)\}=\conv(T)\), we have \(v_G(\nu)<0\) for every \(\nu\in\conv(T)\). The set \(\conv(T)\) is compact and \(v_G\) is continuous, so there is \(\zeta>0\) such that
\[
v_G(\nu)\leq-\zeta\qquad\text{for every }\nu\in\conv(T).
\]
Every distribution over outcomes induced by an expert's prior or posterior belongs to \(\conv(T)\). Since information costs are nonnegative, reporting without learning and every available experiment give the expert a payoff of at most \(-\zeta\). This remains true after taking the supremum over experiments. Hence, no expert can obtain a nonnegative payoff from accepting, and the contract cannot strongly screen charlatans.
\end{proof}

Part \ref{prop1-nolearning} of \cref{prop:general-limits} restates in our setting the impossibility result of \citet{olszewski2007contracts}. A closely related version for restricted sets of outcome distributions appears in \citet{babaioff2011valuable}. Both results ask whether an uninformed forecaster can randomize over reports so as to secure a nonnegative payoff under every possible outcome distribution. To connect this formulation to ours, observe that
\begin{equation}\label{eq:general-minimax}
\max_{x\in X_G}\min_{\tau\in T}\tau\cdot x
=
\max_{x\in X_G}\min_{\nu\in\conv(T)}\nu\cdot x
=
\min_{\nu\in\conv(T)}\max_{x\in X_G}\nu\cdot x
=
\min_{P\in\Delta(T)}v_G(\rho(P)),
\end{equation}
where the first equality follows from linearity, the second from the minimax theorem applied to the bilinear function \((x,\nu)\mapsto\nu\cdot x\), and the last from \(\{\rho(P)\colon P\in\Delta(T)\}=\conv(T)\). The left-hand side is the charlatan's maxmin value, namely, the highest payoff he can guarantee by randomizing over reports, whereas the right-hand side is his lowest value from accepting as his prior \(P\) ranges over \(\Delta(T)\).
Accordingly, the charlatan's maxmin value is negative if and only if he strictly prefers rejection under at least one prior.

\subsection{Learning and Screening}\label{os11}

We now consider an especially simple member of the class of falsification contracts studied by \citet{falsifiability}. The contract, \((u,d)\), consists of a payment \(u>0\) and a penalty \(d>0\). If Bob accepts, he announces an outcome he claims is impossible. If he announces \(\omega_i\) and outcome \(\omega_j\) is subsequently realized, his transfer is
\[
t(\omega_i,\omega_j)=u-d\mathbf 1\{i=j\},
\]
where \(\mathbf{1}\) denotes the indicator function. In words, Bob receives \(u\) unless the outcome he announced is realized, in which case he receives \(u-d\). Given any belief over outcomes, Bob announces an outcome he considers least likely. We show that the expert's ability to learn allows Alice to choose \((u,d)\) that screens charlatans.\footnote{The limits in \cref{prop:general-limits} apply to the broader class of contracts introduced above. The result below establishes sufficiency, but not necessity, of the contract \((u,d)\) for screening.}

If expert Bob accepts and does not acquire information, his expected payoff under prior \(Q\) is
\begin{equation}\label{eq:static-value}
\overline V_{(u,d)}(Q)=\max_{i=1,\ldots,n}\{u-d\mu_i\}=u-d\min_{i=1,\ldots,n}\mu_i.
\end{equation}
If he acquires information using experiment \(E=(S,\chi)\) and reports optimally after observing its signal, his expected contractual payoff is
\[
V_{(u,d)}(E,Q)=\int_S\left(u-d\min_{i=1,\ldots,n}\mu_i^s\right)\dd\overline\chi_Q(s).
\]
The experiment reduces his probability of paying the penalty \(d\) by
\[
Y(E,Q)\coloneqq\min_i\mu_i-\int_S\min_i\mu_i^s
\dd\overline\chi_Q(s).
\]
Because \(\mu\mapsto\min_i\mu_i\) is concave and posterior beliefs over outcomes average to \(\mu\), \(Y(E,Q)\geq0\). His expected payoff after paying for the experiment can then be written as
\[
V_{(u,d)}(E,Q)-C(E,Q)=\overline V_{(u,d)}(Q)+dY(E,Q)-C(E,Q).
\]

For the contract \((u,d)\), a charlatan with prior \(P\) obtains
\(u-d\min_i\rho_i(P)\). Define
\[
\alpha(T)\coloneqq\max_{\mu\in\conv(T)}\min_i\mu_i.
\]
Because \(\{\rho(P)\colon P\in\Delta(T)\}=\conv(T)\), the charlatan's lowest value across priors is
\begin{equation}\label{eq:charlatan-value}
\min_{P\in\Delta(T)}\left\{u-d\min_i\rho_i(P)\right\}=u-d\alpha(T).
\end{equation}
By \eqref{eq:general-minimax}, \(u-d\alpha(T)\) is also the highest payoff the charlatan can guarantee by randomizing over announcements without knowing which model governs the outcome.

To make the charlatan strictly prefer rejection under some prior, Alice must choose \(u<d\alpha(T)\). By \eqref{eq:static-value}, an expert's payoff without learning is decreasing in \(\min_i\mu_i(Q)\), whose largest possible value is \(\alpha(T)\). Hence, acceptance is hardest to secure without learning for experts whose \(\min_i\mu_i(Q)\) is close to \(\alpha(T)\). For \(\eta>0\), let \(\mathcal H_\eta(T)\) collect these priors:
\[
\mathcal H_\eta(T)\coloneqq
\left\{Q\in\Delta(T)\colon \min_i\mu_i(Q)\geq\alpha(T)-\eta\right\}.
\]

We impose the following joint condition on \(\mathcal E\) and \(C\).

\begin{altassumption}{\(A\)}\label{assumption}
There are \(\varepsilon,\eta>0\) and \(K\in\mathbb R_+\) such that, for every nondegenerate \(Q\in\mathcal H_\eta(T)\), there is an experiment \(E_Q\in\mathcal E\) satisfying \(Y(E_Q,Q)\geq\varepsilon\) and \(C(E_Q,Q)\leq K\).
\end{altassumption}

For a fixed set of available experiments \(\mathcal E\) and cost function \(C\), \cref{assumption} requires that every nondegenerate expert prior \(Q\in\mathcal H_\eta(T)\) have access, at cost at most \(K\), to an experiment that reduces the probability of paying the penalty by at least \(\varepsilon\). Crucially, the same bounds must apply to every such prior because Alice offers a single contract without knowing \(Q\).
These common bounds allow her to choose one finite penalty \(d>K/\varepsilon\) such that \(dY(E_Q,Q)>C(E_Q,Q)\) for every nondegenerate \(Q\in\mathcal H_\eta(T)\). The expert may choose a different experiment depending on his prior.

For example, let \(\Omega=\{0,1\}\) and \(T=\{\tau_0,\tau_1\}\), where \(\tau_i\) assigns probability \(3/4\) to outcome \(i\). Suppose an independent pilot study produces a binary signal satisfying \(\chi(i\mid\tau_i)=3/4\). If \(0<\eta<1/8\), then for every \(Q\in\mathcal H_\eta(T)\), Bob announces the outcome opposite the signal and pays the penalty with probability \(3/8\). Without the signal, his probability of paying the penalty is at least \(1/2-\eta\). Hence,
\[
Y(E,Q)\geq\frac18-\eta>0,
\]
and \cref{assumption} holds with \(\varepsilon=1/8-\eta\) whenever the cost of this experiment is uniformly bounded by \(K\).

\begin{theorem}\label{main}
Suppose \cref{assumption} holds. There exists a contract \((u,d)\) that screens charlatans.
\end{theorem}

\begin{proof}
\cref{assumption} implies \(\alpha(T)>0\).\footnote{If
\(\alpha(T)=0\), then every prior and posterior over models assigns zero probability to at least one outcome.
Hence, \(Y(E,Q)=0\) for every experiment \(E\) and prior \(Q\).
Because \(T\) contains at least two models, nondegenerate priors exist, and every such prior belongs to \(\mathcal H_\eta(T)\), contradicting \cref{assumption}.}
Choose \(d>K/\varepsilon\).
Then choose \(\delta>0\) such that
\[
\delta<d\alpha(T)
\qquad\text{and}\qquad
\delta\leq\min\{d\eta,d\varepsilon-K\},
\]
and set \(u=d\alpha(T)-\delta\).
By \eqref{eq:charlatan-value}, the minimum of the charlatan's acceptance value over \(P\in\Delta(T)\) is \(-\delta<0\). Hence, a charlatan strictly rejects under at least one prior.

Fix a nondegenerate expert prior \(Q\). If \(Q\notin\mathcal H_\eta(T)\), then
\[
\overline V_{(u,d)}(Q)>u-d(\alpha(T)-\eta)=d\eta-\delta\geq0,
\]
where the equality follows from the choice \(u=d\alpha(T)-\delta\). Thus, acceptance is optimal without learning.
If \(Q\in\mathcal H_\eta(T)\), \cref{assumption} supplies an experiment \(E_Q\) with \(Y(E_Q,Q)\geq\varepsilon\) and \(C(E_Q,Q)\leq K\). Because \(\min_i\mu_i(Q)\leq\alpha(T)\), the experiment \(E_Q\) delivers the expert a net payoff of at least \[u-d\alpha(T)+d\varepsilon-K =-\delta+d\varepsilon-K\geq0,\]
and so every expert with a nondegenerate prior can obtain a nonnegative payoff from accepting, while a charlatan strictly rejects under at least one prior.
\end{proof}

The intuition behind \cref{main} can be seen from the expert's participation constraint. Without learning, making acceptance optimal for every expert requires \(u\geq d\alpha(T)\), which also makes acceptance optimal for the charlatan under every prior. Under \cref{assumption}, experts for whom acceptance is hardest to secure without learning can improve their payoffs by at least \(d\varepsilon-K\). By choosing \(d>K/\varepsilon\), Alice can therefore set \(u<d\alpha(T)\) while preserving expert participation. The charlatan cannot obtain this improvement and consequently strictly rejects under some priors.

Earlier work resolves the same tension by creating a hole in the set of theories or outcome distributions that the test or contract must accommodate. \citet{olszewski2009strategic} restrict the permissible theories to a particular nonconvex domain and show that informed and uninformed forecasters can be distinguished when the theory governing the data belongs to that domain. \citet{babaioff2011valuable} constrain the possible outcome distributions and specify a common prior over them. They reveal that screening is possible when the prior-averaged outcome distribution does not itself belong to the restricted set: the uninformed forecaster evaluates the contract under the average distribution, while Alice need only make acceptance optimal for informed forecasters whose outcome distributions belong to the restricted set.

\cref{assumption} likewise requires a gap in \(T\). In particular, no distribution attaining \(\alpha(T)\) can belong to \(T\). If such a distribution belonged to \(T\), nondegenerate priors converging to certainty about that model would eventually lie in \(\mathcal H_\eta(T)\), while even full revelation would reduce the probability of paying the penalty by an amount converging to zero. What distinguishes our framework is that the gap is in \(T\), not in \(\conv(T)\), the set of outcome distributions induced by priors over \(T\). For any distribution attaining \(\alpha(T)\), some nondegenerate prior over \(T\) induces it, and Alice must make acceptance optimal for an expert with that prior. Information acquisition allows Alice to do so even when \(u<d\alpha(T)\), while a charlatan with the same prior strictly prefers rejection.

Our construction relies on Bob's risk neutrality and Alice's ability to scale the penalty \(d\) and lump-sum payment \(u\): the gain from learning is \(dY(E,Q)\), while its cost does not scale with \(d\).\footnote{If penalties are capped at \(\overline d\), the construction still works whenever \(\overline d\varepsilon>K\).}

\subsection{Discussion}
 \paragraph{Charlatan priors.} \eqref{eq:general-rejection} characterizes the charlatan's strict-rejection set under any contract \(G\). It shows that \(\mathcal R_G\) depends on \(P\) only through the induced distribution over outcomes \(\rho(P)\); moreover, \(\mathcal R_G\) is convex and open in \(\Delta(T)\).
 If \(G\) screens charlatans, then \(\mathcal R_G\) is nonempty by definition and, by part \ref{prop1-strong} of \cref{prop:general-limits}, is a proper subset of \(\Delta(T)\). For the contract \((u,d)\), \eqref{eq:general-rejection} specializes to
\[
\mathcal R_{u,d}\coloneqq\left\{P\in\Delta(T)\colon \min_i\rho_i(P)>u/d\right\}.
\]
For the contract constructed in the proof of \cref{main}, \(\mathcal R_{u,d}\) is nonempty. Because \(\mathcal R_{u,d}\) is open in \(\Delta(T)\) and nondegenerate priors are dense in \(\Delta(T)\), \(\mathcal R_{u,d}\) contains a nondegenerate prior \(P\).
An expert with prior \(Q=P\) can obtain a nonnegative payoff from accepting, while a charlatan with the same prior strictly prefers rejection. Accordingly, the expert and the charlatan can respond differently while holding the same prior because only the expert can acquire information.

\paragraph{Uniform bounds.} In the proof of \cref{main}, Alice uses \(\varepsilon\) and \(K\) to choose the penalty \(d\), and uses \(\eta\), \(\varepsilon\), and \(K\) to determine how far below \(d\alpha(T)\) she can set \(u\). If the same triple \((\varepsilon,\eta,K)\) satisfies \cref{assumption} for every pair \((\mathcal E,C)\) under consideration, these choices produce one contract that screens charlatans for every pair. If the applicable triple may vary with \((\mathcal E,C)\), however, \cref{assumption} does not ensure that any fixed contract screens charlatans for every pair, as the following example illustrates.

Fix \(\Omega=\{0,1\}\) and \(T=\{\tau_0,\tau_1\}\), where \(\tau_i\) assigns probability \(3/4\) to outcome \(i\). For each
finite \(c \ge 0\),
suppose that the only available experiments are the uninformative experiment and a fully informative experiment that costs \(c\). For every finite \(c\), \cref{assumption} holds with \(\eta=\varepsilon=1/8\) and \(K=c\). Nevertheless, no fixed contract \(G\) screens charlatans for every \(c\). Because \(G\) has a compact message space and continuous transfers, the largest possible gain from full revelation is finite. Once \(c\) exceeds this gain, the expert does not acquire information. Part \ref{prop1-nolearning} of \cref{prop:general-limits} then implies that any contract acceptable to every expert is also acceptable to a charlatan under every prior.

\paragraph{A charlatan who can learn.} The assumption that the charlatan cannot acquire information is stronger than necessary. To establish screening, it is enough to find one prior at which the charlatan rejects. Choose \(P^\ast\in\Delta(T)\) such that \(\min_i\rho_i(P^\ast)=\alpha(T)\), so that the charlatan's probability of paying the penalty before acquiring information is as large as possible. Suppose that no experiment available to him reduces this probability by more than \(\gamma\), where \(0\leq\gamma<\min\{\alpha(T),\eta,\varepsilon\}\). His value from accepting at \(P^\ast\), after accounting for any information acquisition, is therefore at most \(u-d\alpha(T)+d\gamma\). In contrast, under \cref{assumption}, every expert for whom acceptance is hardest to secure without learning can acquire information that raises his payoff by at least \(d\varepsilon-K\). Taking \(d>K/(\varepsilon-\gamma)\) leaves Alice room to choose \(u\) so that acceptance is optimal for every expert while the charlatan strictly prefers rejection at \(P^\ast\).\footnote{Specifically, choose \(u\) such that \(\max\{0,d(\alpha(T)-\eta),d(\alpha(T)-\varepsilon)+K\}<u<d(\alpha(T)-\gamma)\).} In sum, screening does not require the charlatan to be completely unable to acquire information but merely requires a sufficiently large difference between how much the two types can reduce the probability of paying the penalty.\footnote{Simply assuming that the charlatan's experiments are less informative is not enough, because this need not produce a smaller reduction in the probability of paying the penalty.}

\paragraph{Conditional independence of the signal and outcome.} Our model assumes that the signal and outcome are conditionally independent given \(\tau\), as in the independent pilot-study example. Relaxing this assumption permits experiments that provide information directly about the eventual outcome, either alone or together with information about the model.
Part \ref{prop1-nolearning} of \cref{prop:general-limits} is unchanged under this broader formulation, and the proof of \cref{main} goes through with \cref{assumption} stated in terms of posterior distributions over outcomes.
Part \ref{prop1-strong} of \cref{prop:general-limits}
also continues to hold if every expert posterior distribution over outcomes can be induced by some charlatan prior.

\section{Conclusion}\label{conclusion} This paper studies expert testing when an expert's advantage lies in the ability to acquire information and thereby refine his beliefs. When experts cannot acquire information, contracts cannot distinguish experts from charlatans through their willingness to accept. We show that allowing experts to acquire information can overturn this result: a contract can be acceptable to experts yet unacceptable to a charlatan under some beliefs. An expert and a charlatan may even hold the same beliefs yet respond differently because only the expert can acquire further information. However, no contract acceptable to experts can be unacceptable to a charlatan under every belief. We hope this paper encourages further work on distinguishing experts from charlatans when expertise is understood in this way.

\bibliographystyle{jpe}
\bibliography{testing}

@article{alnajjar2010testing,
  title = {Testing Theories with Learnable and Predictive Representations},
  author = {Al-Najjar, Nabil I. and Sandroni, Alvaro and Smorodinsky, Rann and Weinstein, Jonathan},
  journal = {Journal of Economic Theory},
  volume = {145},
  number = {6},
  pages = {2203--2217},
  year = {2010},
  doi = {10.1016/j.jet.2010.07.003}
}

@article{asscal,
  title = {Asymptotic Calibration},
  author = {Foster, Dean P. and Vohra, Rakesh V.},
  journal = {Biometrika},
  volume = {85},
  number = {2},
  pages = {379--390},
  year = {1998}
}

@article{atanasov2020small,
  title = {Small Steps to Accuracy: Incremental Belief Updaters Are Better Forecasters},
  author = {Atanasov, Pavel and Witkowski, Jens and Ungar, Lyle and Mellers, Barbara and Tetlock, Philip},
  journal = {Organizational Behavior and Human Decision Processes},
  volume = {160},
  pages = {19--35},
  year = {2020}
}

@inproceedings{babaioff2011valuable,
  title = {Only Valuable Experts Can Be Valued},
  author = {Babaioff, Moshe and Blumrosen, Liad and Lambert, Nicolas S. and Reingold, Omer},
  booktitle = {Proceedings of the 12th {ACM} Conference on Electronic Commerce},
  pages = {221--222},
  year = {2011},
  doi = {10.1145/1993574.1993608}
}

@unpublished{clark2021contracts,
  title = {Contracts for Acquiring Information},
  author = {Clark, Aubrey and Reggiani, Giovanni},
  note = {Working paper},
  year = {2021}}

@article{dawid1982well,
  title = {The Well-Calibrated {Bayesian}},
  author = {Dawid, A. Philip},
  journal = {Journal of the American Statistical Association},
  volume = {77},
  number = {379},
  pages = {605--610},
  year = {1982}
}

@article{deb2018evaluating,
  title = {Evaluating Strategic Forecasters},
  author = {Deb, Rahul and Pai, Mallesh M. and Said, Maher},
  journal = {American Economic Review},
  volume = {108},
  number = {10},
  pages = {3057--3103},
  year = {2018}
}

@article{dekel2006non,
  title = {Non-{Bayesian} Testing of a Stochastic Prediction},
  author = {Dekel, Eddie and Feinberg, Yossi},
  journal = {The Review of Economic Studies},
  volume = {73},
  number = {4},
  pages = {893--906},
  year = {2006},
  doi = {10.1111/j.1467-937X.2006.00401.x}
}

@unpublished{echenique2007harm,
  title = {You Won't Harm Me If You Fool Me},
  author = {Echenique, Federico and Shmaya, Eran},
  note = {Working paper},
  year = {2007}}

@article{falsifiability,
  title = {Falsifiability},
  author = {Olszewski, Wojciech and Sandroni, Alvaro},
  journal = {American Economic Review},
  volume = {101},
  number = {2},
  pages = {788--818},
  month = {April},
  year = {2011},
  doi = {10.1257/aer.101.2.788}
}

@article{foster1999proof,
  title = {A Proof of Calibration via {Blackwell}'s Approachability Theorem},
  author = {Foster, Dean P.},
  journal = {Games and Economic Behavior},
  volume = {29},
  number = {1--2},
  pages = {73--78},
  year = {1999}
}

@article{gaglianone2022incentive,
  title = {Incentive-Driven Inattention},
  author = {Gaglianone, Wagner Piazza and Giacomini, Raffaella and Issler, Jo{\~a}o Victor and Skreta, Vasiliki},
  journal = {Journal of Econometrics},
  volume = {231},
  number = {1},
  pages = {188--212},
  year = {2022}
}

@incollection{hart2022calibrated,
  title = {Calibrated Forecasts: The Minimax Proof},
  author = {Hart, Sergiu},
  booktitle = {Matching, Dynamics and Games for the Allocation of Resources: Essays in Celebration of {David Gale}'s 100th Birthday},
  editor = {Khan, M. Ali and Sagara, Nobusumi and Zaslavski, Alexander J.},
  series = {Monographs in Mathematical Economics},
  volume = {7},
  pages = {153--159},
  publisher = {Springer Nature Singapore},
  year = {2025},
  doi = {10.1007/978-981-96-6661-4_8}
}

@article{hu2013expressible,
  title = {Expressible Inspections},
  author = {Hu, Tai Wei and Shmaya, Eran},
  journal = {Theoretical Economics},
  volume = {8},
  number = {2},
  pages = {263--280},
  year = {2013},
  doi = {10.3982/TE992}
}

@unpublished{li2026forecasters,
  title = {Incentivizing Forecasters to Learn: Summarized vs. Unrestricted Advice},
  author = {Li, Yingkai and Libgober, Jonathan},
  note = {Working paper},
  year = {2026}}

@article{mellers2015identifying,
  title = {Identifying and Cultivating Superforecasters as a Method of Improving Probabilistic Predictions},
  author = {Mellers, Barbara A. and Stone, Eric and Murray, Terry and Minster, Angela and Rohrbaugh, Nick and Bishop, Michael and Chen, Eva and Baker, Joshua and Hou, Yuan and Horowitz, Michael C. and Ungar, Lyle and Tetlock, Philip E.},
  journal = {Perspectives on Psychological Science},
  volume = {10},
  number = {3},
  pages = {267--281},
  year = {2015}
}

@unpublished{muller,
  title = {Rational Inattention via Ignorance Equivalence},
  author = {M\"{u}ller-Itten, Mich\`{e}le and Armenter, Roc and Stangebye, Zachary},
  note = {Working paper},
  year = {2023}}

@incollection{olssurvey,
  title = {Calibration and Expert Testing},
  author = {Olszewski, Wojciech},
  booktitle = {Handbook of Game Theory with Economic Applications},
  volume = {4},
  pages = {949--984},
  publisher = {Elsevier},
  year = {2015}
}

@article{olszewski2007contracts,
  title = {Contracts and Uncertainty},
  author = {Olszewski, Wojciech and Sandroni, Alvaro},
  journal = {Theoretical Economics},
  volume = {2},
  number = {1},
  pages = {1--13},
  year = {2007}
}

@article{olszewski2008manipulability,
  title = {Manipulability of Future-Independent Tests},
  author = {Olszewski, Wojciech and Sandroni, Alvaro},
  journal = {Econometrica},
  volume = {76},
  number = {6},
  pages = {1437--1466},
  year = {2008},
  doi = {10.3982/ECTA7428}
}

@article{olszewski2009manipulability,
  title = {Manipulability of Comparative Tests},
  author = {Olszewski, Wojciech and Sandroni, Alvaro},
  journal = {Proceedings of the National Academy of Sciences},
  volume = {106},
  number = {13},
  pages = {5029--5034},
  year = {2009},
  doi = {10.1073/pnas.0812602106}
}

@article{olszewski2009nonmanipulable,
  title = {A Nonmanipulable Test},
  author = {Olszewski, Wojciech and Sandroni, Alvaro},
  journal = {The Annals of Statistics},
  volume = {37},
  number = {2},
  pages = {1013--1039},
  year = {2009},
  doi = {10.1214/08-AOS597}
}

@article{olszewski2009strategic,
  title = {Strategic Manipulation of Empirical Tests},
  author = {Olszewski, Wojciech and Sandroni, Alvaro},
  journal = {Mathematics of Operations Research},
  volume = {34},
  number = {1},
  pages = {57--70},
  year = {2009},
  doi = {10.1287/moor.1080.0347}
}

@article{olszewski2011principal,
  title = {The Principal-Agent Approach to Testing Experts},
  author = {Olszewski, Wojciech and P{\k{e}}ski, Marcin},
  journal = {American Economic Journal: Microeconomics},
  volume = {3},
  number = {2},
  pages = {89--113},
  year = {2011},
  doi = {10.1257/mic.3.2.89}
}

@article{pomatto2021testable,
  title = {Testable Forecasts},
  author = {Pomatto, Luciano},
  journal = {Theoretical Economics},
  volume = {16},
  number = {1},
  pages = {129--160},
  year = {2021},
  doi = {10.3982/TE3767}
}

@unpublished{rappoport2017incentivizing,
  title = {Incentivizing Information Design},
  author = {Rappoport, Daniel and Somma, Valentin},
  note = {Working paper},
  year = {2017}}

@article{sandroni2003calibration,
  title = {Calibration with Many Checking Rules},
  author = {Sandroni, Alvaro and Smorodinsky, Rann and Vohra, Rakesh V.},
  journal = {Mathematics of Operations Research},
  volume = {28},
  number = {1},
  pages = {141--153},
  year = {2003}
}

@article{sharma2024procuring,
  title = {Procuring Unverifiable Information},
  author = {Sharma, Salil and Tsakas, Elias and Voorneveld, Mark},
  journal = {Mathematics of Operations Research},
  volume = {50},
  number = {2},
  pages = {1433--1453},
  month = {May},
  year = {2025},
  doi = {10.1287/moor.2022.0085}
}

@unpublished{whitmeyer2023buying,
  title = {Buying Opinions},
  author = {Whitmeyer, Mark and Zhang, Kun},
  note = {Working paper},
  year = {2023}}

@article{yoder2022designing,
  title = {Designing Incentives for Heterogeneous Researchers},
  author = {Yoder, Nathan},
  journal = {Journal of Political Economy},
  volume = {130},
  number = {8},
  pages = {2018--2054},
  year = {2022}
}

@unpublished{zermeno2011,
  title = {A Principal-Expert Model and the Value of Menus},
  author = {Zerme\~{n}o, Luis},
  note = {Working paper},
  year = {2011}}

\end{document}